\documentclass[aps,prl,superscriptaddress,reprint]{revtex4-1}
\usepackage{graphicx}
\usepackage{amsmath}
\usepackage{amssymb}
\usepackage[pdfencoding=auto]{hyperref}

\begin{document}

\title{Modular Berry Connection}

\author{Bart{\l}omiej Czech} 
\affiliation{Institute for Advanced Study, Princeton, NJ, USA}
\author{Lampros Lamprou} 
\affiliation{Massachusetts Institute of Technology, Cambridge, MA, USA}
\author{Samuel McCandlish}
\affiliation{Boston University, Boston, MA, USA}
\author{James Sully}
\affiliation{McGill University, Montr\'{e}al, QC, Canada}
\vskip 0.25cm
\date{\today}

\begin{abstract}
\noindent
The Berry connection describes transformations induced by adiabatically varying Hamiltonians. We study how zero modes of the modular Hamiltonian are affected by varying the region that supplies the modular Hamiltonian. In the vacuum of a 2d CFT, global conformal symmetry singles out a unique modular Berry connection, which we compute directly and in the dual AdS$_3$ picture. In certain cases, Wilson loops of the modular Berry connection compute lengths of curves in AdS$_3$, reproducing the differential entropy formula. Modular Berry transformations can be measured by bulk observers moving with varying accelerations. 
\end{abstract}


\maketitle

\textit{Introduction.}--- 
The last decade of research in the AdS/CFT correspondence \cite{holography} has revealed that quantum entanglement in conformal field theory (CFT) plays a central role in the emergence of the bulk anti-de Sitter (AdS) space-time. On the CFT side, the entanglement structure of a state $\rho$ is encapsulated by the reduced density matrices $\rho_A = {\rm Tr}_{A^c} \rho$ of regions $A$. The most publicized example of how the AdS spacetime geometrizes the CFT entanglement is the Ryu-Takayanagi proposal \cite{rt}, which relates the von Neumann entropy of $\rho_A$ to the area of an extremal surface anchored on the borders of $A$. But other aspects of $\rho_A$ also have crisp gravity duals. The flow generated by $H_{\rm mod}(A) = -\log \rho_A$ (the modular Hamiltonian of $A \subset {\rm CFT}$) acts in the bulk in the same way as the bulk modular Hamiltonian, obtained from the reduced density matrix of perturbative fields in a certain bulk region $W{(A)}$ called the entanglement wedge \cite{jlms}. This fact lies at the core of the recently proven assertion \cite{tomaitor, adh} that knowing $\rho_A$ suffices to reconstruct the perturbative physics in $W{(A)}$.

All reduced density matrices $\rho_A$ descend from the same global state $\rho$. On the gravity side, likewise, the different entanglement wedges $W{\left(A\right)}$ are patches of one global geometry. The common origin of the $\rho_A$s and $W{\left(A\right)}$s suggests a new perspective on entanglement in the global state---one that focuses on the \emph{relations} between different reduced density matrices. The present paper makes a first step in this direction. We study the space of zero modes of the modular Hamiltonians $H_{\rm mod}{\left(A\right)} = -\log \rho_A$, which in the holographic context play a special role because they localize on the respective Ryu-Takayanagi surfaces \cite{tomaitor}. Our objective is to study the maps between them as one takes a region $A$ to another region $A'$.  In more formal language, we are interested in a connection on the bundle of modular zero-modes, fibered over the space of CFT subregions.

That varying the Hamiltonian induces a geometric transformation of Hamiltonian eigenspaces was first understood by Berry in his seminal paper \cite{berry}. We are interested in a new incarnation of Berry's problem wherein the Hamiltonians are modular Hamiltonians (all drawn from the same global state of a CFT) and the parameter space is simply the space of CFT subregions. What defines the notion of parallel transport in the bundle of modular zero modes? What operators have non-trivial monodromies and, in holographic theories, what information do \emph{modular Berry phases} yield about the bulk AdS spacetime? We answer these questions in the setting of the AdS$_3$/CFT$_2$ correspondence, and give a CFT formulation when the global state is $\rho = |0\rangle \langle 0|$.

\textit{A review of the Berry connection.}--- Berry studied a system, which evolves by a slowly varying Hamiltonian $H(\lambda(t))$. Here $t$ is the physical time while $\lambda$ is a coordinate on the parameter space from which the Hamiltonians $H(\lambda(t))$ are drawn. Consider a closed loop in parameter space traversed over a time $t \in [0, T)$, that is $\lambda(0) = \lambda(T)$. If the spectrum of the Hamiltonian is nondegenerate and remains so throughout the trajectory, the evolution brings an eigenstate $|E\rangle = |E(\lambda(0))\rangle$ of the initial Hamiltonian to:
\begin{align}
& \!\!\!\!\!\exp( -i \!\int_0^T E(\lambda(t')) dt' ) \times \exp(i\!\oint \Gamma_\lambda d\lambda)\, |E\rangle~~\textrm{with} 
\label{gentimeevol} \\
& \!\!\!\!\!\Gamma_\lambda = i \langle E(\lambda) \, | d/d\lambda\,  | E(\lambda) \rangle 
\label{berryconn}. 
\end{align}
The Berry phase is the second factor in (\ref{gentimeevol}) and $\Gamma_\lambda$ is the Berry connection.

Equations~(\ref{gentimeevol}-\ref{berryconn}) follow from the adiabatic theorem, which says that if a slowly varying system starts out in a Hamiltonian eigenstate $|E(\lambda(0))\rangle$, it remains in an eigenstate of the instantaneous Hamiltonian. The dynamics is therefore limited to the time dependence of the state's phase. The first term in (\ref{gentimeevol}) accumulates the ordinary dynamical phases $e^{-iE(\lambda(t))dt}$. The second factor in (\ref{gentimeevol}) has a more interesting, geometric origin: it arises from a \emph{precession} of the instantaneous Hamiltonian eigenbasis. 

A Berry transformation around a closed trajectory in parameter space is always valued in a symmetry of an energy eigenspace of $H(\lambda)$. In the simplest setup reviewed above, this symmetry was the freedom of rotating the phase of an energy eigenstate. When additional operators commute with the Hamiltonian, they produce degeneracies in its spectrum and generate symmetries of its energy eigenspaces. In such cases, Berry transformations can rotate the degenerate eigenstates into one another, again acting by automorphisms of energy eigenspaces \cite{zeewilczek}. 

The symmetries emphasized above are local symmetries; they act independently at every point $\lambda$. The local character of the automorphisms of energy eigenspaces of $H(\lambda)$ is the reason why the Berry connection is a gauge connection. To wit, we may transform a given abelian Berry connection to another by
\begin{equation}
\Gamma_\lambda \to \Gamma_\lambda + \partial_\lambda \xi,
\label{eq:berry-connection-ambiguity}
\end{equation}
 where $\xi$ is a continuous function in parameter space. The gauge transformation (\ref{eq:berry-connection-ambiguity}) changes the local eigenbases of $H(\lambda)$ by the phase $e^{-i \xi(\lambda)}$.
  The non-Abelian generalization, where $\xi$ is valued in the Lie algebra of the automorphism group of a given eigenspace of $H(\lambda)$, is well-known. 

\textit{Modular zero modes.}--- Consider a Lorentzian CFT in a global state $\rho$ and study the scenario of adiabatic evolution described above. In the role of the Hamiltonians $H(\lambda)$ we cast $H_{\rm mod}(A)$, the modular Hamiltonians of connected subregions $A$. (From now on, we shall denote CFT subregions with $\lambda$ instead of $A$. This is to emphasize that---from Berry's perspective---choosing a region amounts to choosing a modular Hamiltonian.) However, instead of focusing on Berry phases acquired by modular eigenstates, it is convenient to monitor the monodromies of a special family of operators: modular zero modes $B_i(\lambda)$. These are CFT operators, which satisfy
\begin{equation}
[H_{\text{mod}}(\lambda), B_i(\lambda)] =0.
\label{comm}
\end{equation}
The index $i$ labels distinct zero modes of $H_{\text{mod}}(\lambda)$.

When the CFT has a gravity dual, the operators $B_i(\lambda)$ gain an additional significance. Ref.~\cite{tomaitor} showed that scalar CFT operators which satisfy (\ref{comm}) are holographically dual to bulk operators localized on $[\lambda]$, the extremal (Ryu-Takayanagi) surface anchored on $\lambda$. This is guaranteed by the equivalence between the bulk and boundary modular flows \cite{jlms} and the fact that $[\lambda]$ does not transform under bulk modular flow. 
In holography, modular Berry transformations will therefore reorganize bulk operators localized on Ryu-Takayanagi surfaces.

\begin{figure}[!t]
\begin{center}
\includegraphics[width=0.8\columnwidth]{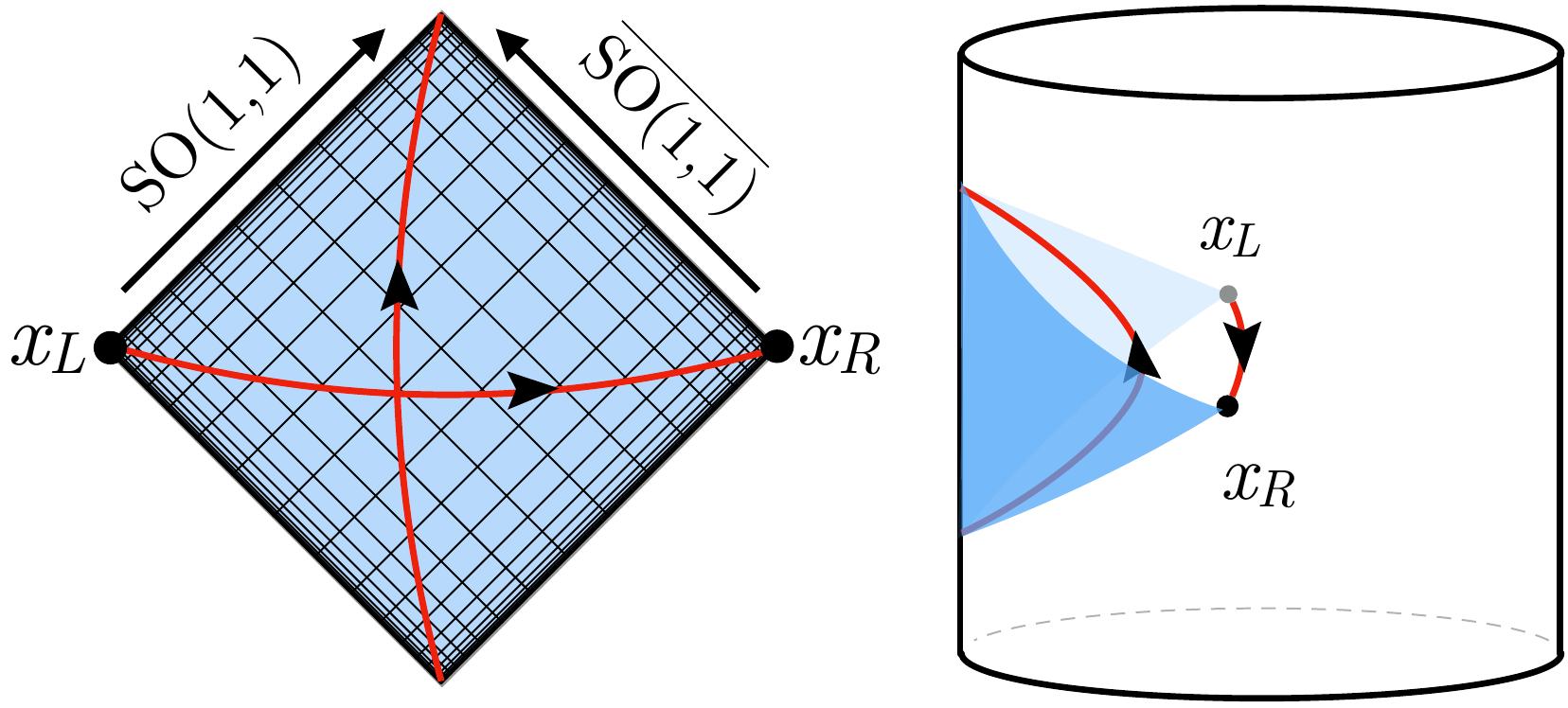}
\caption{A causal diamond in 1+1 dimensions is stabilized by an $\rm{SO}{(1,1)} \times \overline{\rm{SO}{(1,1)}}$ conformal symmetry.  Their symmetric combination is the modular Hamiltonian, which induces a flow from the bottom to the top of the diamond.  The antisymmetric combination induces a flow from the left to the right endpoint. In the bulk of AdS$_3$ these symmetries generate, respectively, trajectories of accelerating observers and translations along spacelike geodesics.}
\label{fig:so11-flow}
\end{center}
\end{figure}

\textit{Modular zero modes in the CFT$_2$ ground state.}---
We concentrate on the ground state $\rho = |0\rangle \langle 0|$ of a 1+1-dimensional CFT. In this case the modular Hamiltonians for the regions considered are all related by conformal transformations.
Conformal symmetry is then sufficient to fix the modular connection, up to a gauge redundancy. 

The connected regions of a CFT$_{2}$ are causal diamonds, labeled by $\lambda$. (It is not useful to distinguish individual spatial slices of the same causal diamond because their reduced density matrices are related by unitary time evolution.) 

A two-dimensional causal diamond is stabilized by an $\rm{SO}(1,1)\times \overline{\rm{SO}(1,1)}$ subgroup of the global conformal group $\rm{SO}(2,2)$. The two factors transform the left and right-moving lightcone coordinates of the diamond; see fig.~\ref{fig:so11-flow}. The symmetric combination of their generators is the vacuum modular Hamiltonian $H_{\text{mod}}(\lambda)$, which generates the flow by modular time. The antisymmetric combination of the two $SO(1,1)$s, which we call $P_D(\lambda)$, implements translations along the modular time-slices. Because the two operators commute
\begin{equation}
[H_{\rm mod}(\lambda), P_D(\lambda)] = 0,
\end{equation}
$P_D(\lambda)$ defines an $SO(1,1)$ symmetry of the modular Hamiltonian and, consequently, of the space of its zero modes. It is therefore convenient to organize the modular zero modes in eigenoperators of this symmetry.

One basis, which spans the zero modes of $H_{\rm mod}(\lambda)$, comprises familiar CFT objects called OPE blocks \cite{stereoscopy}. Intuitively, OPE blocks are an operator basis for the operator product expansion (OPE) of two spacelike-separated local operators:
\begin{align}
\mathcal{O}_{L}{\left(x_{L}\right)}\mathcal{O}_{R}{\left(x_{R}\right)}=\sum_{\Delta} & \left|x_{R}-x_{L}\right|^{-\Delta_{L}-\Delta_{R}}c_{LRi}\times \label{opeexp}\\
 & \times\underbrace{\left|x_{R}-x_{L}\right|^{\Delta}\left(\mathcal{O}_{\Delta}+{\rm descendants}\right)}_{\text{OPE Block } B_\Delta^\kappa(\lambda)}\nonumber
\end{align}
The upper sum in (\ref{opeexp}) runs over the primary operators in the CFT. The correct combination of descendants in the second line of (\ref{opeexp}) is the OPE block. We shall denote OPE blocks $B_\Delta^\kappa(\lambda)$, where $\lambda $ is the causal diamond whose spatial corners are $x_L$ and $x_R$. The label $\kappa$, which parametrizes the zero modes, is related to the external operators in eq.~(\ref{opeexp}) as $\kappa = \Delta_R - \Delta_L$. 

The transformation generated by $H_{\rm{mod}}$ acts locally at $x_L, x_R$ as a boost. When $\mathcal{O}_L, \mathcal{O}_R$ are scalars, $H_{\rm{mod}}$ commutes with the left side of (\ref{opeexp}). This shows that scalar OPE blocks are automatically modular zero modes.

The transformation generated by $P_D$ acts locally at $x_L, x_R$ as a positive and negative dilatation, respectively. Thus, $\mathcal{O}_{L}{\left(x_{L}\right)}\mathcal{O}_{R}{\left(x_{R}\right)}$ is an eigenoperator of $P_D$:
\begin{equation}
[P_D, \mathcal{O}_{L}\mathcal{O}_{R}] = i \kappa\, \mathcal{O}_{L}\mathcal{O}_{R}
\end{equation}
The OPE blocks on the right hand side of (\ref{opeexp}) transform in the same way.  Under a finite transformation of magnitude $s_0$, the OPE blocks transform as
\begin{equation}
B_\Delta^\kappa(\lambda) \to e^{s_0 \kappa} B_\Delta^\kappa(\lambda),
\label{transf}
\end{equation}
so that the effect is a change of normalization.  Eq.~(\ref{transf}) means that it is not possible to specify a $\kappa \neq 0$ OPE block without picking an $SO(1,1)$ gauge or, equivalently, a conformal frame for the causal diamond of interest. This ambiguity in the normalization of OPE blocks is analogous to the phase ambiguity of energy eigenstates in the usual Berry problem. 

\textit{Modular Berry connection in the CFT$_2$ ground state.}---  
Since $B_\Delta^\kappa(\lambda)$s transform under a symmetry of $H_{\rm mod}(\lambda)$, a closed trajectory that visits different causal diamonds will in general bring an OPE block to itself \emph{up to an $SO(1,1)$ transformation}, which is analogous to the second factor in (\ref{gentimeevol}). To compute this transformation, we need an analogue of eq.~(\ref{berryconn}): a connection that maps the zero modes $B_\Delta^\kappa(\lambda)$ to $B_\Delta^\kappa(\lambda+d\lambda)$. As we shall see, up to a choice of gauge, there is a unique connection compatible with conformal symmetry. 

To find it, let us examine the parameter space of our Berry problem. It consists of causal diamonds $\lambda$. Because any causal diamond can be mapped to any other by the global conformal group, and because the state $\rho = |0\rangle \langle 0|$ does not break this symmetry, our parameter space must be a coset space of $SO(2,2)$. Since, as we noted above, a causal diamond is stabilized by an $SO(1,1)\times \overline{SO(1,1)}$ subgroup, the parameter space is:
\begin{equation}
\mathcal{K} = \frac{SO(2,2)}{SO(1,1) \times \overline{SO(1,1)}} =   \frac{SO(2,1)}{SO(1,1)} \times \frac{\overline{SO(2,1)}}{\overline{SO(1,1)}}\,.
\label{ks}
\end{equation}
This object---the space of causal diamonds---was studied in \cite{intgeom, stereoscopy} (see also \cite{janmichalrob}) and named kinematic space. 

Recognizing our parameter space as a coset space highlights a useful fact: that the stabilizer group of a causal diamond $\lambda$ (the group under which the modular zero modes transform) is the isometry group of the tangent space of $\lambda \in \mathcal{K}$. 
From \eqref{transf}, we see that the OPE block transforms under the two $SO(1,1)$ isometries like it is a (possibly non-integer) power of certain combination of vector components:
\begin{align}
\mathrm{Antisymmetric}: \quad  v^{z_L} v^{\bar{z}_L} &\rightarrow e^{s_0}\,\, v^{z_L} v^{\bar{z}_L} \nonumber\\
\mathrm{Symmetric}: \quad  v^{z_L} v^{\bar{z}_L} &\rightarrow v^{z_L} v^{\bar{z}_L}
\label{eq:transformation-construction}
\end{align}
Noting this, we can then immediately write down an appropriate connection for the OPE block. This is because the coset space (\ref{ks}) is a metric space: a product of two two-dimensional de Sitter geometries. Using CFT lightcone coordinates for presenting $x_L = (z_L, \bar{z}_L)$ and $x_R = (z_R, \bar{z}_R)$, the metric is:
\begin{align}
&ds^2= e^{-2S(z_L,z_R)}dz_Ldz_R +e^{-2\bar{S}(\bar{z}_L,\bar{z}_R)}d\bar{z}_Ld\bar{z}_R\,, \nonumber\\
&\text{with}~~S(z_L,z_R) = \log (z_R - z_L)/\epsilon
\label{ksmetric}
\end{align}
and the same function for $\bar{S}(\bar{z}_L,\bar{z}_R)$. For future reference, note that $S(z_L, z_R)$ is proportional to the left-moving contribution to the vacuum entanglement entropy of the causal diamond $(x_L, x_R)$, with $\epsilon$ setting the UV cutoff.

From eq.~(\ref{ksmetric}) we determine the metric-compatible connection $\Gamma$ on kinematic space for the OPE block:
\begin{align}
&\Gamma = -\frac{\partial}{\partial x_L^\mu} S(x_L,x_R)\ dx_L^\mu\,, \label{spinconnection}	
\\
&\text{with}~~S(x_L,x_R) = S(z_L,z_R) + \bar{S}(\bar{z}_L,\bar{z}_R)\nonumber
\end{align}
In each representation of the $SO(1,1)$ gauge fiber, we multiply the connection by the charge $\kappa$.
Because eq.~(\ref{ksmetric}) is fixed by conformal symmetry, so is the connection~(\ref{spinconnection}).
Its curvature two-form $d\Gamma$ is
\begin{equation}
d\Gamma = \frac{\partial^2 S(x_L,x_R)}{\partial x_L^\mu \partial x_R^\nu} dx_L^\mu \wedge dx_R^\nu
\label{modcurvature}
\end{equation}
so the connection is not flat. As a consequence, modular zero modes will pick up nontrivial holonomies under closed trajectories in kinematic space.

\textit{Gauge freedom.}--- 
The connection $\Gamma$ can be shifted by a total derivative $\Gamma \rightarrow \Gamma + d \Lambda(x_L,x_R)$ without affecting its curvature or any Wilson loops constructed from it.  This represents a gauge freedom due to the local frame in which we view eq. \ref{eq:transformation-construction}.
 In particular, it is amusing to consider the gauge choice $\Lambda(x_L, x_R) = - \log\big(\tilde\epsilon(x_L) \tilde\epsilon(x_R) / \epsilon^2 \big)$. Its effect is equivalent to changing the UV cutoff $\epsilon$ to a nonuniform cutoff $\tilde\epsilon$ that regulates the ultraviolet physics at the left and right corners of the causal diamond differently. Thus, the appearance of our gauge freedom may be understood as the local choice of a reference scale $\epsilon$ to set a local UV cutoff. In a theory without a scale, setting a cutoff is a gauge choice.

\textit{Bulk derivation of the modular Berry connection.}---
The above argument may seem a bit formal. Luckily, the holographic picture is more illuminating:   
Here is a bulk argument, which also leads to connection~(\ref{spinconnection}).

\begin{figure}[!t]
	\begin{center}
		\includegraphics[width=1.0\columnwidth]{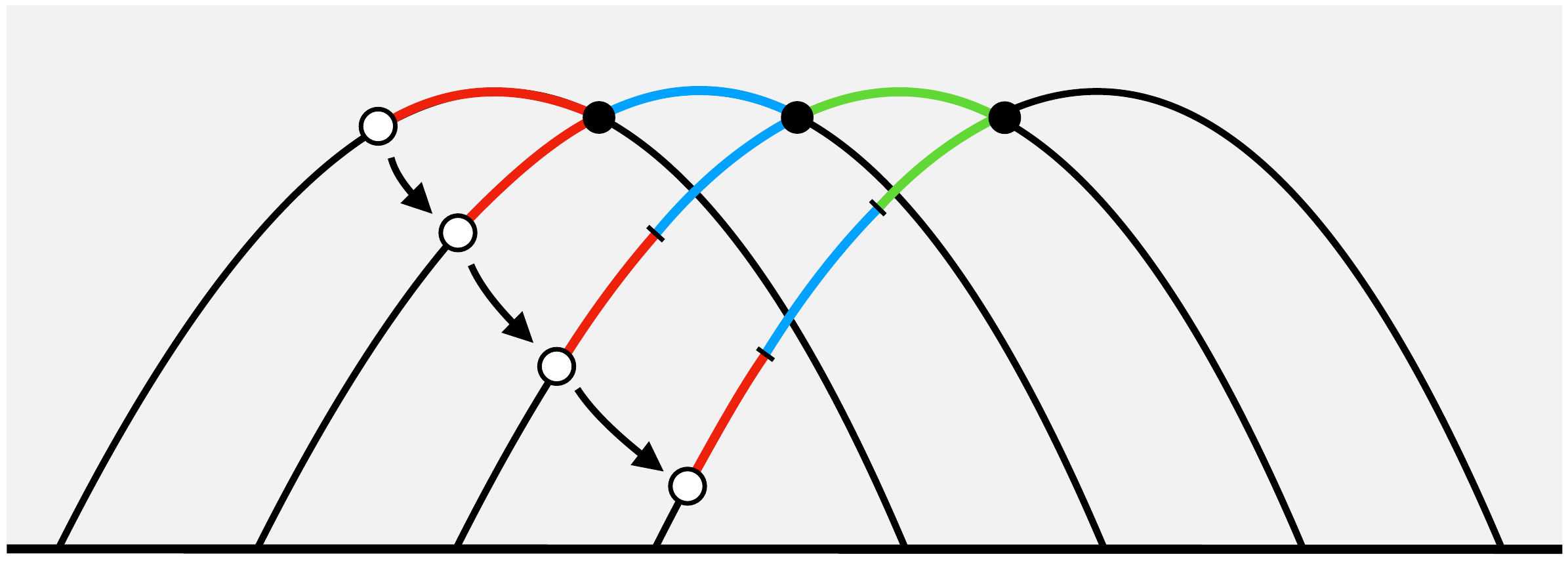}
		\caption{The bulk realization of the modular connection is determined by a map between neighboring geodesics. Given a fixed 1-parameter family of intersecting geodesics $[\lambda(\sigma)]$, a natural gauge is one where the connection leaves the point of intersection between $[\lambda(\sigma)]$ and $[\lambda(\sigma+d\sigma)$] fixed. The origin $s=0$ will then recede as it traverses the family of geodesics. The total precession is the length of the curve connecting all of the intersection points.  We can think of the connection as a ``rolling without slipping'' condition, where the point on the geodesic tangent to this curve is always momentarily at rest.}
		\label{fig:rolling-condition}
	\end{center}
\end{figure}

We describe a connection in the bulk for nearby geodesics $[\lambda]$ and $[\lambda+d\lambda]$ that intersect. (This restriction can be lifted, see \cite{hholes}.)  
The connection is the unique isometric map that takes points on $[\lambda]$ to points on $[\lambda+d\lambda]$ equidistant to the intersection point (see fig. \ref{fig:rolling-condition}). 
Here we have implicitly identified points by their distance to the intersection point. This identification is really a choice of gauge. 
A general gauge varies by an $SO(1,1)$ translation that will label points with respect to different origins on each geodesic.

To see the effect of the connection on operators, note that a scalar OPE block is holographically dual to the bulk operator \cite{stereoscopy,daCunha:2016crm}:
\begin{equation}
B_\Delta^\kappa(\lambda) = N \int_{[\lambda]} ds\, \phi_\Delta(s)\, e^{-\kappa s}
\label{bulkopeblock}
\end{equation}
Here $s$ is the proper length parameter along the bulk geodesic $[\lambda]$ (in units of $L_{\rm AdS}$) and $\phi_\Delta$ is the bulk operator dual to $\mathcal{O}_\Delta$. The $SO(1,1)$ generated by $P_D(\lambda)$ comprises translations along $[\lambda]$; see fig.~\ref{fig:so11-flow}. A translation by $s_0$ induces a shift $s \to s-s_0$, which transforms the right hand side of (\ref{bulkopeblock}) as in eq.~(\ref{transf}). Once again, it is not possible to present an OPE block without picking an $SO(1,1)$ gauge. As noted above, choosing the gauge amounts to selecting a privileged point on the geodesic, which serves as the origin of the local coordinate $s$ on $[\lambda]$.

Now consider a closed trajectory $\lambda(\sigma)$ in kinematic space such that the geodesics $[\lambda(\sigma)]$ and $[\lambda(\sigma + d\sigma)]$ intersect for all $\sigma$.  
Let us inspect what happens to some privileged point $s=s_0$ under a kinematic trajectory. Our connection maps the point $s=s_0$ on $[\lambda(\sigma)]$ to one equidistant from the intersection point on $[\lambda(\sigma+d\sigma)]$. Transforming the point through a sequence of intersections results in a precession of the point $s=s_0$, which recedes farther and farther away from successive intersection points. This is illustrated in fig.~\ref{fig:rolling-condition}. After the trajectory closes, the amount by which the privileged point $s=s_0$ recedes equals the sum of the distances between consecutive intersection points. In a continuous limit, this distance becomes the circumference of the bulk curve that sits on the common envelope of the geodesics $[\lambda(\sigma)]$.

Ref.~\cite{diffent} gave a useful formula for the circumference of the envelope of a sequence of geodesics:
\begin{equation}
	\ell = -\oint \frac{\partial S(x_L,x_R)}{\partial x_L^\mu} dx_L^\mu 
\end{equation}
The quantity $S(x_L, x_R)$ is the same as in eq.~(\ref{spinconnection}). 
This is (the logarithm of) the Wilson loop of connection (\ref{spinconnection}), so our bulk definition of the connection agrees with the boundary computation.
The normalization of a scalar OPE block $B_\Delta^\kappa$ changes under this trajectory by:
\begin{equation}
B_{\Delta}^\kappa (\lambda) \rightarrow e^{\kappa \cdot \ell}\, B_{\Delta}^\kappa (\lambda). \label{sdiff}
\end{equation}

When all $[\lambda(\sigma)]$ live on a static slice of the bulk geometry, eq.~(\ref{sdiff}) can be rewritten as a two-dimensional integral over all geodesics in the static slice, which intersect the bulk curve \cite{intgeom}. The integrand of that formula, called the Crofton form, can be thought of as a density of geodesics. We now recognize the Crofton form as the curvature two-form of the modular Berry connection (\ref{modcurvature}). 

\textit{A bulk measurement of the Berry transformation.}--- Consider Alice who travels through AdS$_3$ with uniform acceleration. Assume the acceleration is sufficient to produce a Rindler horizon, which is a spacelike geodesic $[\lambda]$. Alice's time evolution is generated by $H_{\rm mod}(\lambda)$, i.e. the proper time along the trajectory is proportional to modular time. While the physics behind $[\lambda]$ remains hidden behind the Rindler horizon, the value of the field $\phi_\Delta$ at a point $s$ on $[\lambda]$ is marginally accessible to Alice. To measure it, she couples her detector to the modular zero mode $\phi_\Delta(s)$. This modular zero mode can be obtained from the OPE blocks $B_\Delta^\kappa(\lambda)$ studied in this paper by an inverse Laplace transform with respect to $\kappa$. 

Our modular Wilson loops describe how zero modes transform due to changes of $\lambda$. To measure such a Wilson loop, Alice will have to conduct a Berry experiment where the modular Hamiltonian that generates her evolution changes adiabatically in time. She can achieve this by controling the magnitude and direction of her instantaneous acceleration so that her time translations are generated by the modular Hamiltonians of a sequence of Rindler wedges. For a closed trajectory in kinematic space, Alice would start and end by evolving according to the same $H_{\rm mod}(\lambda)$. Note that Alice has an infinite amount of modular time to fiddle with her acceleration and explore different Hamiltonians along the way, so that any kinematic trajectory can be traversed adiabatically. As is usual in measurements of Berry transformations, at the end of her journey Alice can compare notes with a friend who moved with unchanging acceleration. If they initially coordinated their labeling of fields $\phi_\Delta(s)$ on $[\lambda]$, Alice will find her labels shifted by $\ell$ from eq.~(\ref{sdiff}). 
 
\textit{Extensions.}--- It is an intriguing problem to define a physically motivated modular Berry connection in states other than the vacuum. In holographic theories we could repeat the argument in fig.~\ref{fig:rolling-condition} for a connection that maps geodesics isometrically, but the CFT meaning of this definition is unclear. In some applications, we may consider connections valued in geodesic reparameterizations---the full symmetry group of the geodesic. 

We hope that the concept of modular Wilson loops will prove useful in understanding the construction of local bulk operators \cite{kll} because modern treatments \cite{tomaitor, klmod} emphasize the importance of modular Hamiltonians. We also hope that modular Wilson loops will offer a new light on holographic complexity \cite{complexity}. At least in the vacuum, the concept also makes sense away from holographic settings, so it may be useful for studying critical systems.

\begin{acknowledgments}
\textit{Acknowledgments.}--- We thank Herman Verlinde for a thought-provoking conversation about Berry phases and complexity, Xiaoliang Qi for the suggestion to think about coset spaces of the conformal group, Daniel Harlow and Jieqiang Wu for conversations at the early stages of this work. We also thank Vijay Balasubramanian, Alex Belin, Jan de Boer, Alejandra Castro, Diego Hofman, Jorrit Kruthoff, Aitor Lewkowycz, Sagar Lokhande, Matt Headrick, Juan Maldacena, Alex Maloney, Mark Van Raamsdonk, Moshe Rozali, Erik Verlinde, Guifr{\'e} Vidal and Edward Witten for valuable discussions. BC would like to thank  
the University of Amsterdam and UT Austin for hospitality and an opportunity to present this work at an early stage. The work of BC is supported by the NSF grant NSF PHY-1314311 and funds from the Institute for Advanced Study and Swarthmore College. The work of LL is supported by the Pappalardo Fellowship. The work of SM is supported by the Simons Collaboration Grant on the Non-Perturbative Bootstrap. The work of JS is supported by the Natural Sciences and Engineering Research Council of Canada (NSERC) Banting Postdoctoral Fellowships program. This research was supported in part by the National Science Foundation under Grant No. NSF PHY-1125915.
\end{acknowledgments}

\end{document}